\documentclass[aps,prl,twocolumn,showpacs]{revtex4}
\usepackage{amssymb}
\usepackage{amsmath}
\usepackage{epsfig}

\begin{document}

\title{Long-Range Effects in Layered Spin Structures}
\author {Alessandro Campa${}^{1}$, Ramaz Khomeriki${}^{2,3}$,  David Mukamel${}^4$, Stefano Ruffo${}^2$}
\affiliation{${\ }^{(1)}$Complex Systems and Theoretical Physics Unit, Health and Technology Department, \\
Istituto Superiore di Sanit\`a, and INFN Roma1 Gruppo Collegato Sanit\`a, Viale Regina Elena 299, 00161
Roma, Italy \\
${\ }^{(2)}$Dipartimento di Energetica ``S. Stecco" and CSDC,
Universit\`a di Firenze and INFN, via s. Marta, 3, 50139 Firenze, Italy\\
${\ }^{(3)}$  Department of Exact and Natural Sciences, Tbilisi State University, 0128 Tbilisi, Georgia \\
${\ }^{(4)}$Department of Physics of Complex Systems, Weizmann Institute of Science, Rehovot 76100, Israel}

\begin{abstract}
We study theoretically layered spin systems where long-range
dipolar interactions play a relevant role. By choosing a specific
sample shape, we are able to reduce the complex Hamiltonian of the
system to that of a much simpler coupled rotator model with
short-range and mean-field interactions. This latter model has
been studied in the past because of its interesting dynamical and
statistical properties related to exotic features of long-range
interactions. It is suggested that experiments could be conducted
such that within a specific temperature range the presence of
long-range interactions crucially affect the behavior of the
system.
\end{abstract}

\maketitle

Many of the forces that we see in the universe have a long-range
nature where in $d$ dimensions a pairwise interaction potential
decays as $V(r)\sim 1/r^{d+s}$ with $-d\le s\le 0$. Examples
include gravitational interactions, Coulomb and magnetic forces.
Nowadays there is a vast literature which describes various exotic
nonlinear and statistical properties of many-body systems with
long-range interactions such as disagreement of predictions of
microcanonical and canonical ensembles, negative specific heat and
temperature jumps, etc.~\cite{ruffobook}. However, most of these
studies are purely theoretical and there have been only few
suggestions on how to test experimentally all these peculiarities.
One example is the astrophysical observation of negative specific
heat \cite{astro} which is an outcome of the truly long range
nature of the gravitational force. In addition, it has been
suggested that systems composed of a small number of particles can
show negative specific heat \cite{part} due to non-additivity even
when the interaction is short range. This has been verified
experimentally in nuclear collisions \cite{nuclear}, atomic sodium
clusters \cite{sodium} and in molecular clusters \cite{cluster}.
However no experimental test of these predictions has been carried
out for a laboratory system with long range interactions.
\begin{figure}[t]
\hspace{-1cm} \epsfig{file=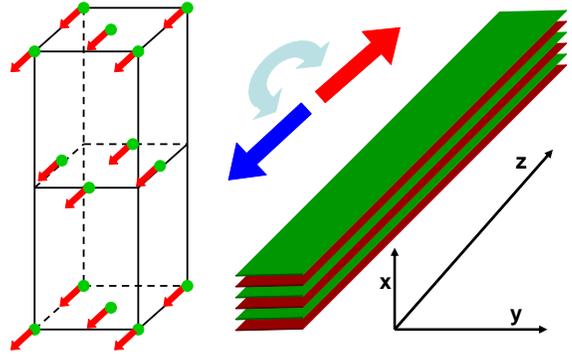,width=0.8\linewidth}
\caption{Left graph: schematic arrangement of spin $s=1/2$
$Cu^{+2}$ ions in a $(C_\nu H_{2\nu+1}NH_3)_2CuCl_4$ layered
compound. Right graph: Suggested form of the sample which allows
the observation of the spontaneous flips of the magnetization
vector (thick arrows at the middle) which are predicted by
Hamiltonian (\ref{rotor}) to which the microscopic model (\ref{1})
effectively reduces.} \label{xyz}
\end{figure}

This Letter aims at proposing testable effects of dipolar magnetic
forces, whose long-range nature is a consequence of the cubic
decay law of the potential $V(r) \sim 1/r^3$ . This law results in
a strong dependence of the dipolar energy on the sample shape (see
e.g. Ref. \cite{landau}). However, in ordinary magnetic systems,
dipolar energy is about a thousand times smaller than Heisenberg
exchange interactions between nearby spins. Therefore, in most
cases the role of long-range forces is to introduce some
anisotropy which determines the ordering direction in the magnetic
sample. On the other hand, in nuclear magnets, where magnetic
order is fully defined by dipolar interaction, one has to go down
to nano Kelvin temperatures to observe ordering \cite{oja}.

In this Letter, we propose to examine more closely magnetic
layered spin structures (e.g. $(C_\nu H_{2\nu+1}NH_3)_2CuCl_4$
\cite{sievers})  in which the effective magnetic interaction
between electronic spins is predominantly dipolar. In particular,
we examine rod shaped layered spin structures, whose microscopic
Hamiltonian can be effectively reduced to that of a one
dimensional coupled rotator model with both nearest neighbor
coupling and a dominant mean-field interaction term resulting from
the dipolar forces. We suggest that these compounds provide a
system in which exotic phenomena that characterize coupled rotator
models with both short and long-range couplings
\cite{campa,mukruf} could readily be observed. Here, we propose to
verify experimentally the presence of all-to-all mean-field
couplings by monitoring the time-scales of spontaneous
magnetization flips below the magnetic transition. Similar effects
of magnetization reversals could in principle take place in
systems with only short-range interactions \cite{ising}. However,
here we propose to explore the experimental conditions in which
collective reversals are driven by the presence of long-range
interactions, and therefore their existence becomes strongly
dependent on the shape of the sample. Moreover, the average
reversal time is expected to grow as the exponential of the volume
of the sample as opposed to the case of short range interactions
where it grows only as the exponential of the surface area
\cite{ising}. In addition, magnetization reversals due to long
range interactions are expected to be sharper, since they involve
at once the entire sample.

The magnetic arrangement of the class of compounds $(C_\nu
H_{2\nu+1}NH_3)_2CuCl_4$ is schematically given in Fig. 1 (see for
more details Refs. \cite{dupas,miedema}). The system is composed
of ferromagnetic layers with strong intralayer interactions, $W$,
and a weak coupling, $w$, between the layers which is either
ferromagnetic for $\nu=1$ or antiferromagnetic for $\nu>1$. This
allows us to consider the three-dimensional spin system at
temperatures or energies well below $W$ as a quasi-one dimensional
ferromagnetic or anti-ferromagnetic spin chain consisting of
classical spins, with each spin in the chain representing a whole
layer. The forces in the system are provided by the exchange
interactions between the effective spins in the chain (short-range
forces) and the dipolar interactions among all spins (long-range
forces). The interlayer exchange interaction turns out to be
comparable with the dipolar interactions making these systems
excellent candidates for considering long-range effects, even in
the case of a small number of layers. Note that, as expected, the
shape of the sample is rather important in systems with dipolar
interactions. A particularly interesting case is that of
ferromagnetic interlayer coupling ($\nu=1$) with a sample shape
for which dipolar forces favor ferromagnetic ordering such as in a
rod cut along the layer planes (rod along axis $z$ in Fig. 1).
This case will be discussed in detail below.

These systems have been modelled previously including only the
anisotropic contribution of the dipolar forces (see e.g. Ref.
\cite{dupas}) while the long-range character of these forces was
neglected. At this stage we include fully the dipolar interactions
and the Hamiltonian reads
\begin{widetext}
\begin{equation}
{\cal H}=-W\sum\limits_{I,<i,j>}\left(s_{Ii}^zs_{Ij}^z+\eta s_{Ii}^ys_{Ij}^y+\xi
s_{Ii}^xs_{Ij}^x\right)-w\sum\limits_{I,<i,j>}\vec s_{Ii}\vec s_{I+1,j}
+\sum\limits_{Ii\neq Jj}\frac{2\mu_B^2}{r^3}\left(\vec s_{Ii}\vec s_{Jj} -3\frac{(\vec
s_{Ii}\vec r)(\vec s_{Jj}\vec r)}{r^2}\right)
\label{1}
\end{equation}
\end{widetext}
where the first sum represents the intralayer exchange interaction
(indices $i$ and $j$ refer to nearest neighbor spins within the
same layer, while the indices $I$ and $J$ number the layers),
$\mu_B$ is Bohr's magneton and $\vec r$ is the vector between
the sites of the spins $\vec s_{Ii}$ and $\vec s_{Jj}$. The parameters $\xi$
and $\eta$ ($<1$) yield biaxial hard axes anisotropies along the
$x$ and $y$ axes, respectively. They are a result of the
crystallographic forces and take into account out-of plane and
in-plane anisotropies (note that the main contribution to the
in-plane anisotropy comes from dipolar forces themselves, see
below). The second sum stands for the interlayer exchange
interactions (with $i$ and $j$ now referring to nearest neighbor
spins in adjacent layers) and the last term describes dipolar
interactions among all spins. According to \cite{dupas,miedema}
$W\simeq 10^4 w$ in the compounds $(C_1 H_{3}NH_3)_2CuCl_4$. At
low temperatures all spins in a single layer are ordered
ferromagnetically and therefore the spin vector $\vec s_{Ii}$
could be considered as independent of the index $i$, $\vec
s_{I}\equiv \vec s_{Ii}$, and this represents the spin of a whole
layer. This is justified only if one works below the ordering
temperature of a single layer, which approximately coincides with
the intralayer exchange constant $W$. Under the additional
condition $nw<W$ ($n$ is the number of spins in a single layer), in
a certain temperature range each layer will be ordered
ferromagnetically, while the transition to 3D ordering will be
strongly affected by the existence of long-range dipolar forces
(which are comparable with the short range interlayer exchange
$w$). We consider the thermodynamic properties of the system under
such conditions. Applying the ordinary procedure to calculate the
dipolar sum, one can divide it into a short-range contribution
(restricting ourselves to consider the interaction between nearest
neighbors) and  a long-range one \cite{landau}. Then \eqref{1} can
be rewritten in the following one-dimensional representation (see
Ref.~\cite{sieversprb})
\begin{widetext}
\begin{equation}
{\cal H} =n\left[B_x\sum\limits_{J=1}^{N} \bigl(s_J^x\bigr)^2+ B_y\sum\limits_{J=1}^{N} \bigl(s_J^y\bigr)^2-
2\omega_{ex}\sum\limits_{J=1}^{N-1} \bigl(s_J^ys_{J+1}^y+s_J^zs_{J+1}^z\bigr)- \frac{\omega_M}{N}
\Bigl(\sum\limits_{J=1}^{N} s_J^z\Bigr)^2+ \frac{\omega_M}{2N}\Bigl(\sum\limits_{J=1}^{N} s_J^y\Bigr)^2\right], \label{2}
\end{equation}
\end{widetext}
where $B_x=4W(1-\xi)$ and $B_y=4W(1-\eta)$ define the hard axis
anisotropies along $x$ and $y$, respectively (the out-of plane
anisotropy being much larger than the in-plane one $B_x\gg B_y$),
$N$ is the number of layers, $\omega_{ex}$ is an effective exchange
constant between the layers consisting of the sum of the exchange
constant $w$ between the neighboring spins of the different layers
and the contribution of dipolar forces between the same spins.
Thus
\begin{equation}
\omega_{ex}=2w-\frac{2\mu_B^2}{r_a^3}\left(2-\frac{3q_a^2}{r_a^2}\right),
\end{equation}
Where $q_a$ and $r_a$ are the distances between nearest
neighboring spins within the same layers and different layers,
respectively. In the case of the considered compound
$q_a=5.25\cdot 10^{-8}$cm and $r_a=9.97\cdot 10^{-8}$cm
\cite{miedema}. Finally the last two mean-field terms come from
the long-range part of dipolar forces. Moreover, $\omega_M=(4\pi/3)(2\mu_B^2/v_0)$ and $v_0$ is the
volume of the unit cell of the lattice. Note that the nearest
neighbor part of dipolar forces favors antiferromagnetic ordering
of the spins in neighboring layers, while the long-range component
of the forces is ferromagnetic along $z$. The measured value for the out-of plane
anisotropy is $B_x=240$mK, while the in-plane anisotropy constant
$B_y$ is less than $5$mK. The exchange constants is
$\omega_{ex}=4.8$mK (see e.g. Ref. \cite{dupas}), and
$\omega_M=16$mK. Thus, we have:
\begin{equation}
B_x/\omega_M=15, \quad B_y/\omega_M\approx 0.2, \quad \omega_{ex}/\omega_M=0.3. \label{constants}
\end{equation}
As is evident, the out-of plane anisotropic term is associated
with a much larger energy scale than all the others and therefore
we have neglected in \eqref{2} all the terms which include the
$s_x$ component of the spins which emerge from dipolar or exchange
interlayer forces.

We now consider the torque equation
\begin{equation}
\frac{d {\vec s_J}}{dt}={\vec s_J} \times {\vec H}_J~,
\label{torque}
\end{equation}
where ${\vec H}_J=-{\partial{\cal H}}/{\partial{\vec s_J}}$ is an
effective magnetic field acting on the spin $\vec s_J$. Noting
that $|\vec s_J|=1/2$, it is natural to adopt the definitions
$\vec S_J\equiv 2\vec s_J$, $S_J^y=\sqrt{1-(S_J^x)^2}\sin\theta_J$
and $S_J^z=\sqrt{1-(S_J^x)^2}\cos\theta_J$ which yield the
equations of motion in terms of angular variable $\theta_J$ and of
$S_J^x$. Observing that $B_x$ is the largest energy scale in the
Hamiltonian , one can simplify the equations of motion for $S_J^y$
as follows
\begin{equation}
\frac{d \theta_J}{dt}=-nB_xS_J^x
\label{sy}
\end{equation}
and approximate $\sqrt{1-(S_J^x)^2}\approx 1$. Then, substituting
$S_J^x$ from the last equation into Hamiltonian \eqref{2}, we
obtain the following effective Hamiltonian, which includes only the
angles $\theta_J$'s, and their time derivatives
\begin{widetext}
\begin{equation}
{\cal H^\prime}=\frac{2{\cal H}}{n\omega_M}= \frac{1}{2}\sum\limits_{J=1}^{N}
\left[\frac{d\theta_J}{dt}\right]^2-\frac{\omega_{ex}}{\omega_M}\sum\limits_{J=1}^{N-1}
\cos(\theta_{J+1}-\theta_J) +\frac{B_y}{2\omega_M}\sum\limits_{J=1}^{N}\sin^2\theta_{J}
-\frac{1}{2N}\Bigl(\sum\limits_{J=1}^{N} \cos\theta_J\Bigr)^2
+\frac{1}{4N}\Bigl(\sum\limits_{J=1}^{N} \sin\theta_J\Bigr)^2~,
\label{rotor}
\end{equation}
\end{widetext}
where the dimensionless time is introduced via the transformation
$t\rightarrow tn\sqrt{B_x\omega_M}$. The above model describes a
system of coupled rotators with nearest neighbor and all-to-all
mean-field interactions.

In analogy with the similar system considered in
Ref.~\cite{campa}, we can derive the thermodynamic and dynamical
properties of the model (\ref{rotor}). It follows that, in the
energy range $-0.3<\varepsilon<\varepsilon_c=0.376$, two maximal
entropy states with opposite magnetization
$m=(\sum\limits_{J=1}^{N} \cos\theta_J)/N$ are present and, in
principle, for finite $N$, the system can flip between these
states. Above the critical energy $\varepsilon_c$ the system is
characterized by a single maximum entropy state at $m=0$. At
$\varepsilon_c$ a second order phase transition of the mean-field
type is present. The lower energy limit derives from the fact
that, in the thermodynamic limit ($N\rightarrow\infty$), the
energy $\varepsilon$ satisfies the condition $\varepsilon>-m^2/2
-\omega_{ex}/\omega_M$. It follows that below
$\varepsilon=-{\omega_{ex}}/{\omega_M}\simeq-0.3$ there is an
ergodicity breaking region \cite{mukruf}, where the system retains
its magnetization direction for infinitely long time. If the
system energy is slightly above the ergodicity breaking limit,
magnetization flips are very rare. Simulating the system using
torque equation \eqref{torque} does not reveal any flip for
observable time scales, i.e. the system behaves like in the
ergodicity breaking region.

Examining the dynamics of the one-dimensional spin chain
\eqref{2} at $\varepsilon=0.2$, sufficiently above the
ergodicity breaking region, the time evolution of
the magnetization displays the expected behavior (see upper graph in Fig.
2). We then compute numerically the probability distribution function (PDF) of
$m$, $P(m)$, and present the data of $(1/N)\log(P(m))$ as circles in the inset.
These are compared (modulo a vertical normalization shift) with the
analytical curve for the entropy derived for Hamiltonian \eqref{rotor}.
The agreement is impressively good, considering the numerous approximations
we have made when deriving Hamiltonian \eqref{rotor} from \eqref{2}.

If now one further increases the system energy above the critical
one, $\varepsilon_c$, magnetization fluctuates around zero and the
corresponding PDF is presented in the lower graph of Fig. 2,
together with the analytical prediction, again in very good
agreement.

\begin{figure}[t]
\centerline{\epsfig{file=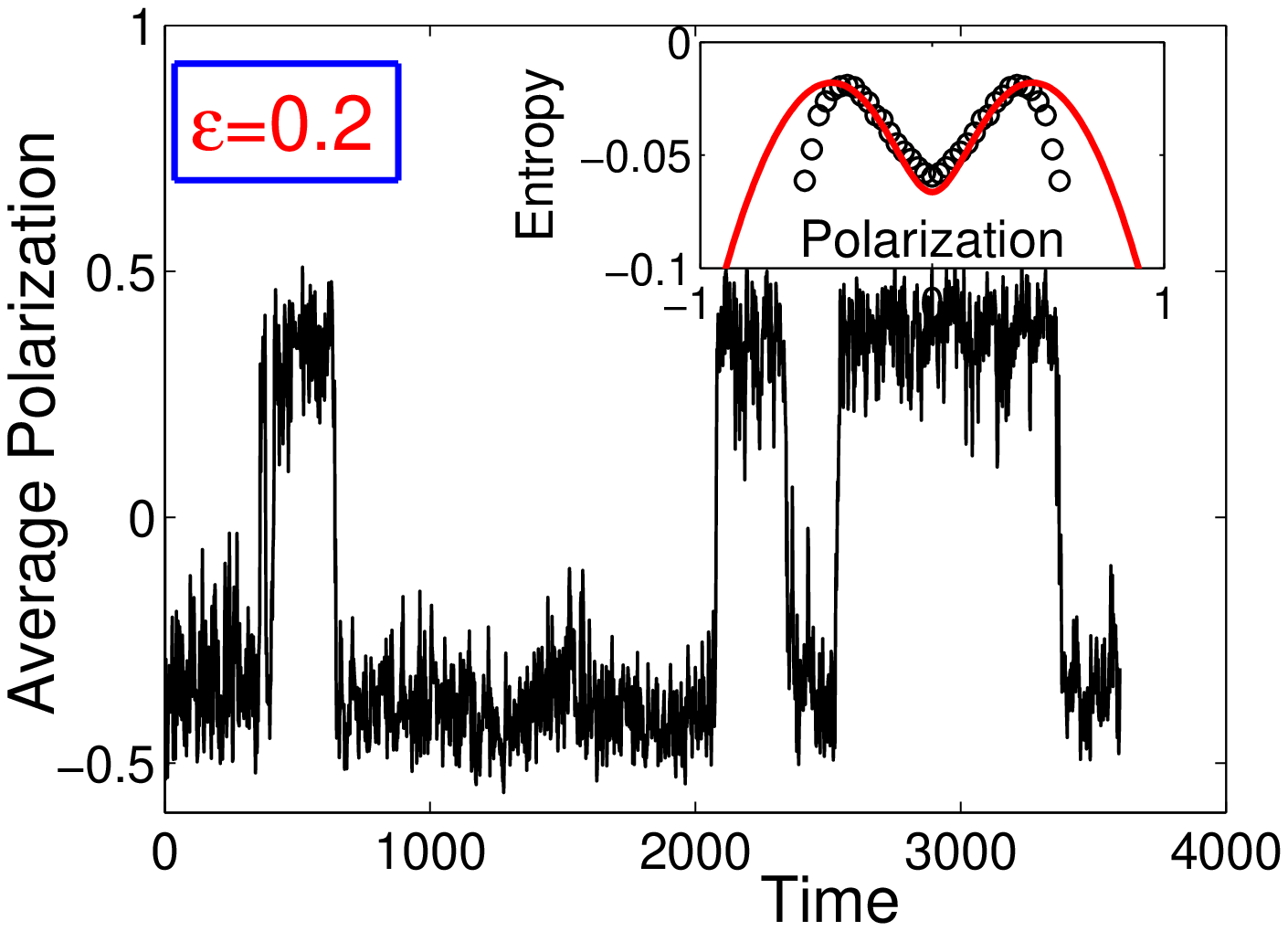,width=0.8\linewidth}}
\epsfig{file=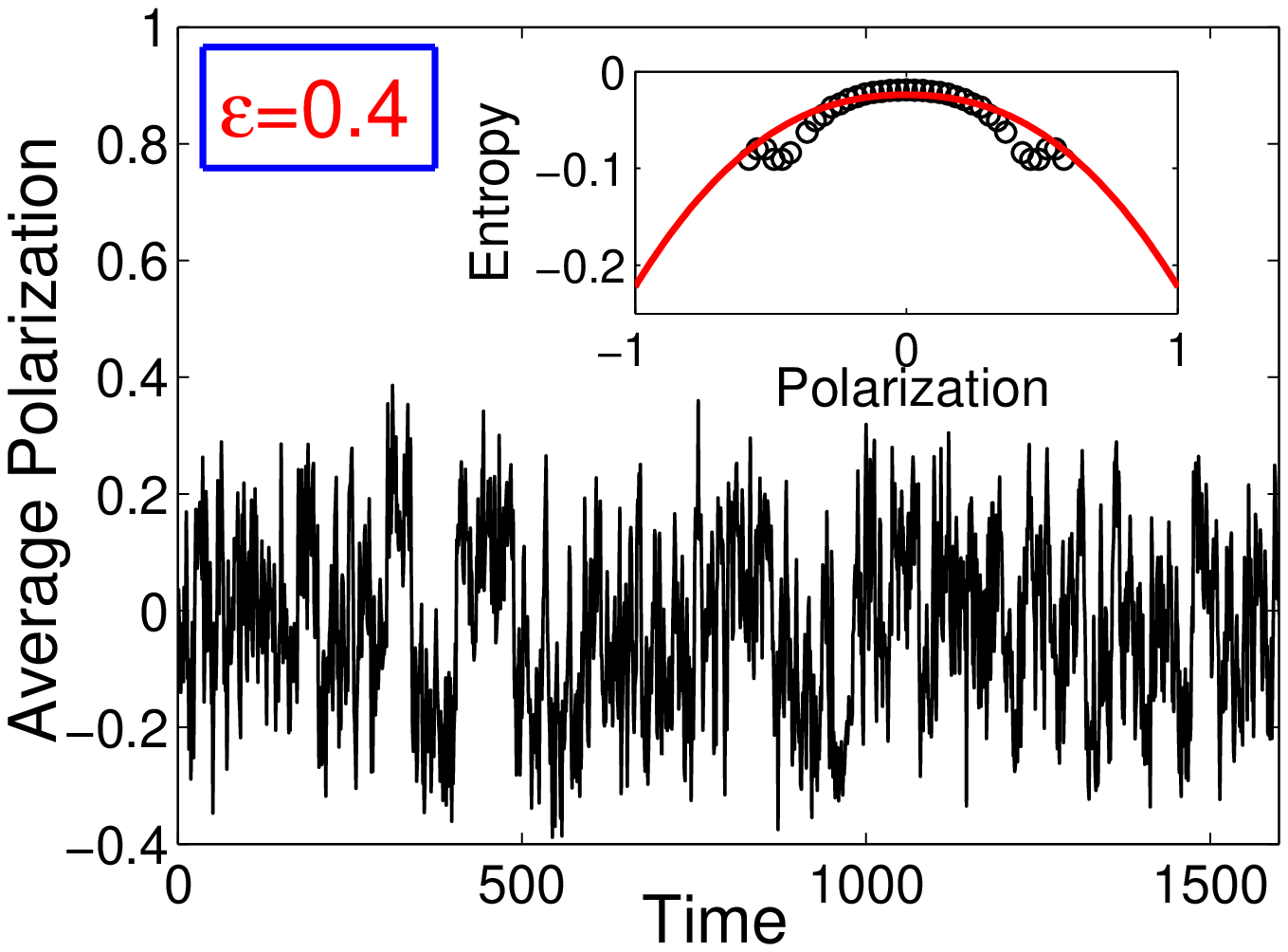,width=0.8\linewidth}
\caption{Main plots: Numerical simulations of the one dimensional spin chain model \eqref{2},
\eqref{torque} with different energies:
$\varepsilon=0.2$ (upper graph) and $\varepsilon=0.4$ (lower graph), i.e. below and
above the second order phase transition energy $\varepsilon_c=0.376$ predicted for
Hamiltonian \eqref{rotor}.
The number of layers in the simulations is taken $N=100$.
Insets show the corresponding entropy curves (solid lines) and data obtained from the PDF of
the magnetization (open circles).}
\label{flips}
\end{figure}

Let us conclude with a few remarks that are relevant for the experimental
implementation. Let us recall that, throughout this paper, we have considered a rod shaped
sample. Moreover, as stated above, the reduction of the
Hamiltonian to a one-dimensional rotator model holds only if $nw<W$. Thus,
for the compound under consideration, $n$ should be less than
$10000$. Hence, in a real experiment, one should examine samples of
typical size $n=20\times500$ within a single layer and, e.g., $N=20$
layers in total. In contrast, in spherical samples with about the
same number of spins per layer, long-range effects will be
negligible and only short-range forces are left in the effective
one-dimensional model: no phase transition will be present and, therefore,
magnetization flips will be absent.

In summary, we have studied theoretically the layered spin
structure $(C_\nu H_{2\nu+1}NH_3)_2CuCl_4$ and showed that, under
certain conditions, it could be modeled as a system of coupled
rotators with both short and mean-field interactions. In
experiments, this could be confirmed by the presence of
magnetization flips for $\nu=1$ (ferromagnetic interlayer
short-range interactions). When $\nu>1$ the short-range interlayer
coupling is antiferromagnetic, while each layer is still ordered
ferromagnetically. For an appropriate choice of the short versus
long-range coupling (which could be precisely controlled by
changing the length of rod shaped sample), the latter compounds
could serve as a laboratory system for which various exotic
phenomena, such as ensemble inequivalence, negative specific heat,
temperature jumps, etc. \cite{campa,new}, characterizing systems
with long-range interactions, could be observed.

We would like to thank A.J. Sievers for multiple useful
suggestions and discussions. R. Kh. acknowledges support by
Marie-Curie international incoming fellowship award (contract  No
MIF1-CT-2005-021328) and USA CRDF Award \# GEP2-2848-TB-06. We
also acknowledge financial support of the Israel Science Foundation
(ISF) and of the PRIN05 grant on {\it Dynamics and
thermodynamics of systems with long-range interactions}. We thank
the Newton Institute in Cambridge (UK) for the kind hospitality
during the programme ``Principles of the Dynamics of
Non-Equilibrium Systems" where part of this work was carried out.

\end{document}